\documentclass[aip,jcp,amsmath,amssymb,reprint,superscriptaddress]{revtex4-1}
\usepackage{graphicx}%
\usepackage{dcolumn}%
\usepackage{bm}%
\usepackage{epstopdf}
\usepackage[hyperindex,breaklinks]{hyperref}
\usepackage[caption=false]{subfig}

\begin{document}

\title{Estimates of bond length and thermal expansion coefficients from x-ray scattering experimental data using reverse Monte Carlo simulations}
\date{\today}
\author{R. Ashcraft}
\affiliation{Department of Physics, Washington University in St. Louis, St. Louis, Missouri, 63130, USA}
\author{K. F. Kelton}
\email[Author to whom correspondence should be addressed.]{kfk@wustl.edu}
\affiliation{Department of Physics, Washington University in St. Louis, St. Louis, Missouri, 63130, USA}
\affiliation{Institute of Materials Science and Engineering, Washington University in St. Louis, St. Louis, Missouri, 63130, USA}
\begin{abstract}
The previously discussed anomalous behavior (i.e. negative) of the thermal expansion coefficient obtained from the pair correlation function is examined in the context of the nearest-neighbor distance (bond length) distribution. The bond length distribution is obtained from a Voronoi tessellation analysis of the atomic structures obtained from both reverse Monte Carlo simulations of x-ray scattering data and molecular dynamics simulations. When a robust measure of central tendency (mean or median) is used a positive thermal expansion is obtained from the temperature-dependent bond length that has the same magnitude as that obtained from direct measurements of the volume as a function of temperature.  The same is true when larger neighbor distances, as obtained in higher order peaks in the pair distribution function are tracked.   This calls into question the recent claim that fragility of metallic liquids is embedded in these higher order peaks.  It also shows that the previously reported anomalous contraction of the bond length arise from tracking the mode, which does not account for the skewness of the distribution.
\end{abstract}

\maketitle

\section{Introduction}\label{intro}
	The static structure factor, $S(q)$, and the related pair distribution function, $g(r)$, obtained from experimental scattering data are routinely used to deduce the linear thermal expansion coefficient, 
\begin{equation}
\beta=\frac{1}{3V}\frac{\mathrm{d}V}{\mathrm{d}T}~\mathrm{,}
\end{equation}
where $V$ is the volume, for crystalline systems by tracking the position of the first peak as a function of temperature.  Following the same method, some studies in metallic liquids have shown an anomalous contraction of the first peak in $g(r)$ with increasing temperature, indicating a negative thermal expansion coefficient. However, values of $\beta$ obtained from direct measurements of the volume have been positive~\cite{Gangopadhyay2014d,Lou2013}. To explain this difference, it was suggested that the coordination number decreased with increasing temperature, forming stronger bonds between the atoms and a decrease in the atomic separation. However, later studies determined that the contraction contraction was likely due to a failure to account for the asymmetry of the nearest-neighbor distance (NND) distribution~\cite{Ding2014,Sukhomlinov2017,Ding2017a,Gangopadhyay2018}. This asymmetry is a consequence of the redistribution of neighboring atoms to typically larger distances due to the anharmonicity of the interatomic potential. It has also been suggested that due to the complex interplay of the partial pair correlation functions it is unlikely that reliable data for $\beta$ can be obtained from the $g(r)$ for the liquid.
	
	One of these studies~\cite{Sukhomlinov2017} suggested a promising approach using a skew normal distribution (SND) to fit the first peak of the total radial distribution function, 
\begin{equation}
R(r)=\frac{g(r)}{4 \pi r^2 \rho}~\mathrm{,}
\end{equation}
where $\rho$ is the number density. This gives an effective nearest neighbor distribution that accounts for the increasing skewness, bypassing many of the issues arising with the use of peak positions. From the fit the mean bond length can be identified and tracked with temperature to obtain an approximate value for $\beta$. However, as the authors point out this approach is not without flaws. For multi-component alloys the main peak in $g(r)$ will contain multiple partial pair correlation functions that may not be well described by a single SND. One option would be to fit each partial $g(r)$ with a SND and then sum them with the usual weighting factors (e.g. Faber-Ziman\cite{Waseda1980} coefficients) to obtain the total $g(r)$. This would then give an effective total NND distribution. Experimentally, however, it is typically difficult to obtain the needed information on all the pair correlation functions to perform this type of analysis.
	
	Here a more detailed examination of the NND distribution is presented that is based on RMC and MD modeling using a Voronoi tessellation. The main conclusion is that unlike $g(r)$ the robust measures of central tendency for the NND distribution are well behaved, exhibiting only expansion, and give reliable information about the linear thermal expansion coefficient. Furthermore, it is shown that the rate of expansion obtained from the NND distribution is equal to the rate obtained from direct volume measurements. This calls into question a recent proposal that liquid fragility~\cite{Angell1995a} (a measure of the deviation of the temperature dependence of the activation energy of the viscosity) is related to the temperature dependence of the peak positions of the 3rd and 4th peak positions in $g(r)$~\cite{Wei2015}. Finally, the results presented here give more evidence of the local nature of fragility, which was recently reported~\cite{Pueblo2017}.

\section{Experimental and Analysis Methods}\label{methods}
\subsection{Experimental Methods}
Equilibrium and supercooled liquid structural data was obtained at the Advanced Photon Source at Argonne National Laboratory on beamline 6ID-D using the beamline electrostatic levitation (BESL) technique. Density and thermal expansion measurements were made from video images of levitated samples back-lit by a high-intensity LED. The details of these experimental methods can be found elsewhere~\cite{Mauro2011b,Chung1996,Gangopadhyay2005,Gangopadhyay2014a,Gangopadhyay2014d}. 

\subsection{Molecular Dynamics Simulations}
Molecular dynamics (MD) simulations were made for 10 compositions using the LAMMPS~\cite{Plimpton1995a} software. Some simulations employed the GPU package of LAMMPS~\cite{Brown2011,Brown2012,Brown2013}. Each simulation consisted of 15,000 atoms contained in a cubic box with periodic boundary conditions in the NPT ($P=0$) ensemble. The Nos\'e-Hoover thermostat was used to equilibrate each system at each temperature before data collection. Ten configurations were used to obtain statistics on the NND distribution.

\subsection{Reverse Monte Carlo Simulations}
Reverse Monte Carlo simulations were made for 34 metallic liquids using the RMC++~\cite{Gereben2007a} software. Structure factors, $S(q)$, obtained from X-ray scattering experiments were used as the only constraint. Reverse Monte Carlo simulations perform a minimization of the $\chi^2$ given by
\begin{equation}
\chi^2 = \sum_{i} \frac{[S'(q)-S(q)]}{\sigma^2}
\end{equation}
where $S'(q)$ is the structure factor calculated from the RMC atomic configuration, $S(q)$ is the experimental data, and $\sigma$ is the reliability of the data set. This minimization is achieved by moving atoms, which are chosen randomly, in a random direction and computing the new $\chi^2$ from this configuration. If the $\chi^2$ is reduced the move is accepted while if the $\chi^2$ increases the move is accepted according to a Boltzmann probability. This procedure is then repeated until the $\chi^2$ is minimized.

	The reliability of such minimally constrained simulation has been examined in a recent study. To generate sufficient statistics and determine error in the Voronoi tessellation each liquid was simulated seven times. Each RMC simulation consisted of 10,000 randomly generated atoms in a cubic box whose size is consistent with the experimental number density of the liquid, using periodic boundary conditions. Minimum cutoff distances and swapping positions between atoms of different elements were used to improve the convergence time. Convergence was assumed when both the magnitude and the change with time of the $\chi^2$ were sufficiently small. 

\subsection{Voronoi Tessellation}
	Voronoi tessellation was performed on each RMC configuration using a Python extension of the Voro++\cite{Rycroft2006,Rycroft2009} software. The Voronoi tessellation procedure can lead to significant errors when systems with different sized atoms are considered~\cite{Park2012}. To account for this radical Voronoi tessellation~\cite{Gellatly1982,Gerstein1995} was used for all systems containing multiple elements. In this method, the distance to each bisecting plane is weighted by the radius of each atom; the Goldschmidt radii~\cite{Gale2004} were used for this weighting. Another error common to Voronoi tessellation is the occurrence of exceptionally small faces and edges compared to the polyhedron average, which occur from slightly more distant atoms~\cite{Brostow1998}. These more distant atoms tend to exaggerate the skewness of the "true" nearest-neighbor distribution. To attempt to account for this effect, small-faces were removed using a percentage of the system polyhedron average face area as the cutoff. Repeating this removal for multiple values gives a determination of the reliability of the final results.
	
	A more robust method for removing more distant nearest neighbors was developed using Gaussian mixtures modeling (GMM), which is a fuzzy clustering algorithm. This method assumes that the NND and face area data are composed of two Gaussian distributions coming from only the more distant "artifact" atoms and the "true" nearest neighbors. The data is then clustered into groups according to the probability of inclusion in each distribution. Using the Scikit-learn Python library~\cite{Pedregosa2012} a single Voronoi tessellation was fit. This fit was then used to predict which distribution each NND and face area pair belong to for the remaining Voronoi tessellations. This method, though more reasonable than a strict cutoff using face area, is still only approximate. Other clustering algorithms that are not mode-seeking and that allow for different size and covariance of clusters could also be used.

\section{Results and Discussions}\label{results}
\begin{figure}
\includegraphics[width=\linewidth]{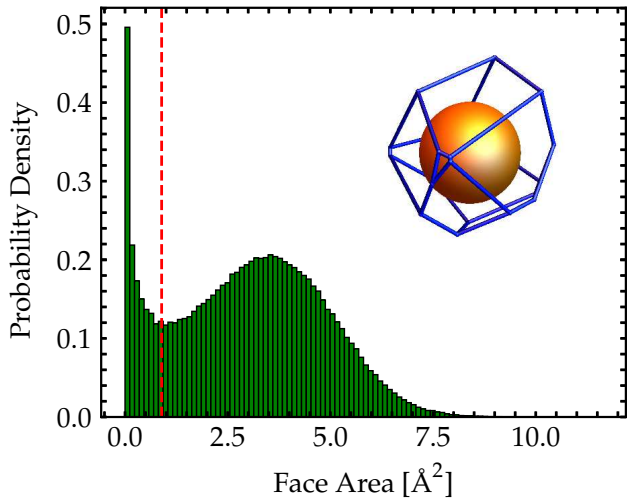}
\caption{\label{fig:Face_Dist} Representative probability density for polyhedron face area for $\mathrm{Zr}_{80}\mathrm{Pt}_{20}$ at $T=1191\mathrm{K}$. A schematic illustration of a Voronoi polyhedron that contains a small face is shown. The dashed line indicates an area cutoff using a fraction (here $0.3$) of the average polyhedral face area.}
\end{figure}

As mentioned in the preceding section a Voronoi tessellation often creates polyhedra with abnormally small faces. Figure~\ref{fig:Face_Dist} shows a typical distribution of the area of the faces of a Voronoi polyhedron and a typical Voronoi polyhedron with a small face. Two distinct features are observed in the distribution, a peak corresponding to the larger-size faces representative of the typical NND and one corresponding to smaller faces, which come from more distant atoms. Although a clear minimum between the two peaks is observed here, providing an obvious choice for a cutoff, this is not always the case. Even for cases where there is an appreciable separation between the two peaks, the minimum becomes less prominent as the temperature of the system increases. Atoms are able to sample smaller and larger distances more frequently consistent with the anharmonicity of the interatomic potential. Since the overlap of these two distributions is significant, separating the two distributions becomes non-trivial. A typical method for determining which faces to remove is to use a fraction of the average polyhedron face area, however the cutoff value from this method is arbitrary. 

\begin{figure}
\includegraphics[width=\linewidth]{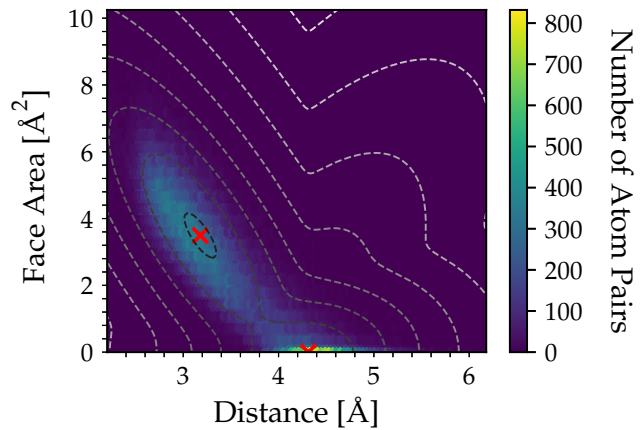}
\caption{\label{fig:GMM_Plot} Representative hexbin plot for polyhedral face areas and nearest-neighbor distances for $\mathrm{Zr}_{80}\mathrm{Pt}_{20}$ at $T=1191\mathrm{K}$. The 'x' marks the centers of each Gaussian cluster. The dashed lines indicate the log-likelihood probability of being from a given distribution (dark shading indicates a high likelihood).}
\end{figure}

\begin{figure}
\includegraphics[width=\linewidth]{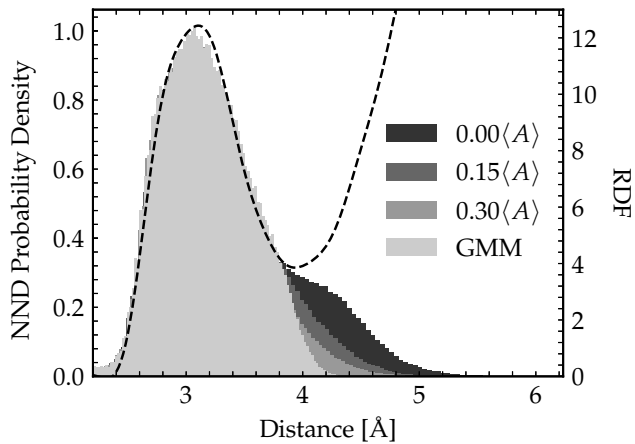}
\caption{\label{fig:NND} Plot of a representative, $\mathrm{Zr}_{80}\mathrm{Pt}_{20}$ nearest-neighbor distance distribution at $T=1191\mathrm{K}$ (left) using each cutoff method (shading darkest to lightest): removing $A<0.0$, $A<0.15 \langle  A \rangle$, $A<0.30 \langle  A \rangle$, (where $A$ is the polyhedron face area) and using the Gaussian mixtures modeling. The radial distribution function (dashed line, right axis) is also shown for comparison.}
\end{figure}

	A less subjective method to separate the distributions (mentioned previously) is to use GMM, in which one Gaussian is centered on the "artifact" atoms and the other represents the true distribution of NNDs and face areas. Figure~\ref{fig:GMM_Plot} shows a representative plot using the GMM method for the case of only two cluster centers and two features. The dashed lines are log-likelihood contour curves showing the probability of inclusion in each probability distribution and the crosses mark the centers of the distributions. This method does a good job of separating the distributions, but there are some limitations. By using a mixture of Gaussians the underlying NND distribution that is deduced is assumed to be Gaussian. However, this distribution is known to be inherently asymmetric due to the anharmonicity of the interatomic potential. It is also clear that the data points do not recreate the log-likelihood curves exactly, again indicating that a Gaussian distribution is not the best assumption. Finally, this method tends to underestimate the skewness since the prediction uses a simple maximum probability to ascertain inclusion in a cluster, meaning that farther atoms will not be included.   By including more features (i.e. polyhedron face perimeter etc.) in the GMM it might be possible to better determine the true NND distribution. However, care needs to be employed when increasing the number of features for multi-component systems, since the different atom types may cause unexpected clustering. In this case it may be necessary to isolate elements even for RMC simulations where chemical effects are not reproduced well. A representative NND distribution is shown in Fig.~\ref{fig:NND} using each of the examined cutoff methods.
	
\begin{figure}
\includegraphics[width=\linewidth]{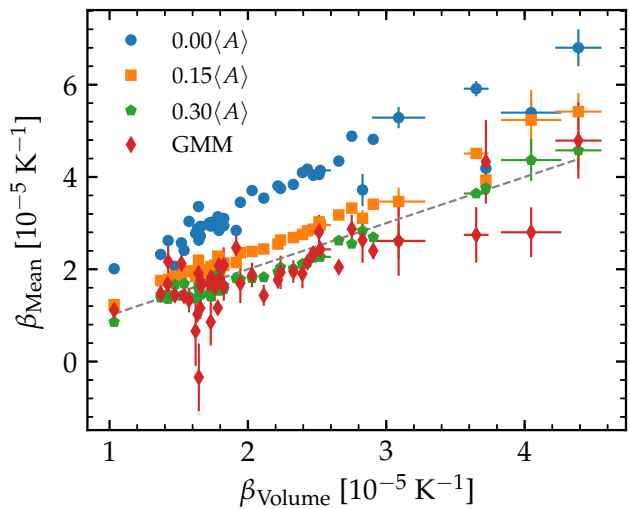}
\caption{\label{fig:Exp_Comp} A comparison of the linear expansion coefficient from the mean of the nearest-neighbor distance distribution ($\beta_{\mathrm{Mean}}$) with that obtained from direct measurements of the volume ($\beta_{\mathrm{Volume}}$) evaluated at the liquidus temperature. The values for $\beta_{\mathrm{Mean}}$ obtained for removing no faces (circle), faces where $A<0.15\langle A \rangle$ (square), faces where $A<0.30\langle A \rangle$ (pentagon) and faces using the Gaussian mixture modeling (diamond). The dashed line shows the case of $\beta_{\mathrm{Mean}}=\beta_{\mathrm{Volume}}$ as a guide for the eye.}
\end{figure}
	
	The linear expansion coefficients estimated from the GMM and fractional cutoff analyses,$\beta_{\mathrm{Mean}}$, are shown as a function of the linear expansion coefficient measured from the volume change, $\beta_{\mathrm{Volume}}$ in Fig.~\ref{fig:Exp_Comp}. A strong correlation between these expansion coefficients exists, regardless of the method used to remove small faces, and even for the case when no faces were removed. This indicates, in contrast to recent claims\cite{Gangopadhyay2018,Sukhomlinov2017}, that the expansivity can be deduced from the structural data with proper modeling. No more information is required other than that used to obtain $g(r)$. It is also important to note that the slope of the best fit line approaches unity if a sufficient number of faces are removed, indicating that the rate of expansion of the first shell is commensurate with that of the bulk. Since the local neighbor configurations expand as the same rate as the volume expansion, it is reasonable to conclude that all higher order coordination shells will expand at this rate as well.
	
	From an examination of the peak positions in $G(r)=4 \pi r \rho [g(r)-1]$ a recent study concluded that the fourth coordination shell is correlated with kinetic fragility~\cite{Wei2015}. Rather than tracking $g(r)$, $G(r)$ was examined since less information is required to obtain that quantity. However, $G(r)$ still exhibits the same asymmetric skewing as $g(r)$, raising doubts about the reliability of the conclusions drawn. In light of the results presented here, the lack of correlation with coordination shells lower than the fourth, especially the first coordination shell, is likely a result of not tracking the NNDs in the proper way. It is not surprising to find a correlation with higher order shells when using these weaker metrics for the central tendency, since the distribution of atoms tends to be less skewed for higher order coordination shells, since atoms are distributed more symmetrically (at longer distances the interatomic potential is more symmetric). If each coordination shell expands at the same rate then all of the metrics, $m_{str}^{(V_{i-j})}$ and $m_{str}^{(r_i)}$ used in the previous study reduce to the volume change between $T_g$ (the glass transition temperature) and $0\mathrm{K}$ or 1/3 of this value, respectively, extrapolated from the liquid which is  a statement that the expansivity is related to the fragility as in~\cite{Gangopadhyay2017}.
	 
	 The expansion coefficients for the best case shown in Fig.~\ref{fig:Exp_Comp}, i.e. removing faces with area less than 30\% of the average face area, are listed in Table~\ref{tab:data}. The linear expansion coefficient calculated from the median and mode of the NND distribution are also listed. Since both the median and the mean use the entire distribution to give a measure of the central tendency it is not surprising that they give consistent estimates of an expansion, while the mode can give both expansion and contraction. The mode then tracks the behavior of the peak position of $g(r)$, technically $R(r)$ the radial distribution function. This analysis supports the conclusion of others~\cite{Ding2014,Gangopadhyay2018} who claim that the anomalous contraction previously reported fails to consider the increased skewness and deviation from Gaussian behavior. In particular, we see that it is dangerous to infer changes in the local configuration using the peak position for either $g(r)$ or $R(r)$, since they are not consistent estimators of the underlying atomic distribution.

\begin{table*}
\caption{\label{tab:data} Data for the linear expansion coefficient, $\beta$, for each composition using the mean, median, and mode of the nearest-neighbor distance distribution (removing faces with area $A <0.3\langle A \rangle$), where $\langle A \rangle$ is the average face area, and the value from volumetric measurements evaluated at the liquidus temperature (melt for elementals and phase diagram values for MD systems which were not previously calculated). Error estimates are listed in parentheses. ($^\dagger$) denotes compositions which were simulated using MD.}
\begin{ruledtabular}
\begin{tabular}{p{0.19\linewidth}>{\centering}p{0.13\linewidth}>{\centering}*{4}{p{0.17\linewidth}}}
\hline
Composition & Liquidus ($\mathrm{K}$) &      $\beta_{\mathrm{Mean}}$ ($10^{-5} \mathrm{K}^{-1}$) &    $\beta_{\mathrm{Median}}$ ($10^{-5} \mathrm{K}^{-1}$) &      $\beta_{\mathrm{Mode}}$ ($10^{-5} \mathrm{K}^{-1}$) &  $\beta$ ($10^{-5} \mathrm{K}^{-1}$) \\
\hline
$\mathrm{Al}$                                                                    &      933 &      3.6(0.1) &      3.1(0.1) &     -3.0(1.0) &      3.65(0.09) \\
$^\dagger\mathrm{Al}$(~\cite{Mendelev2008})                                      &      926 &      4.6(0.4) &      4.0(0.3) &     0.61(0.2) &        4.4(0.2) \\
$^\dagger\mathrm{Cu}$(~\cite{Mendelev2008})                                      &     1353 &      2.6(0.4) &      2.0(0.2) &     0.41(0.1) &        3.1(0.2) \\
$\mathrm{Cu}_{30}\mathrm{Zr}_{30}\mathrm{Ti}_{40}$                               &     1113 &    1.78(0.03) &     1.5(0.04) &     0.37(0.9) &    1.945(0.001) \\
$\mathrm{Cu}_{46}\mathrm{Zr}_{54}$                                               &     1198 &    1.94(0.02) &    1.61(0.02) &     0.15(0.5) &  2.2169(0.0007) \\
$\mathrm{Cu}_{47}\mathrm{Zr}_{47}\mathrm{Al}_{6}$                                &     1180 &    2.04(0.03) &    1.84(0.03) &      2.0(2.0) &    2.236(0.001) \\
$\mathrm{Cu}_{50}\mathrm{Zr}_{42.5}\mathrm{Ti}_{7.5}$                            &     1152 &    2.14(0.02) &    2.02(0.01) &      6.0(1.0) &     2.43(0.001) \\
$\mathrm{Cu}_{50}\mathrm{Zr}_{45}\mathrm{Al}_{5}$                                &     1173 &  2.117(0.008) &    2.07(0.03) &      7.0(1.0) &    2.394(0.001) \\
$\mathrm{Cu}_{50}\mathrm{Zr}_{50}$                                               &     1222 &    2.01(0.02) &     1.8(0.04) &      3.0(1.0) &    2.328(0.002) \\
$^\dagger\mathrm{Cu}_{50}\mathrm{Zr}_{50}$(~\cite{Mendelev2009})                 &     1222 &    1.79(0.02) &    1.63(0.02) &      1.8(0.3) &      1.78(0.01) \\
$\mathrm{Cu}_{60}\mathrm{Zr}_{20}\mathrm{Ti}_{20}$                               &     1127 &    2.55(0.02) &    2.44(0.02) &    0.069(0.3) &    2.751(0.001) \\
$\mathrm{Cu}_{64}\mathrm{Zr}_{36}$                                               &     1200 &    2.69(0.01) &     2.9(0.01) &    0.086(0.3) &    2.906(0.003) \\
$\mathrm{Ge}$                                                                    &   1211   &    3.76(0.03) &    3.87(0.03) &      1.8(0.6) &     3.72(0.002) \\
$\mathrm{LM601}$                                                                 &     1157 &    1.83(0.02) &     1.4(0.05) &    -0.33(0.8) &    2.115(0.002) \\
$\mathrm{Ni}$                                                                    &     1728 &    2.84(0.06) &      2.4(0.2) &      2.0(2.0) &  2.82865(3e-05) \\
$^\dagger\mathrm{Ni}$(~\cite{Mendelev2012})                                      &     1728 &      4.4(0.5) &      3.4(0.3) &     0.28(0.2) &        4.0(0.2) \\
$\mathrm{Ni}_{59.5}\mathrm{Nb}_{40.5}$                                           &     1448 &    1.81(0.04) &    1.75(0.05) &     0.39(0.3) &     2.03(0.002) \\
$^\dagger\mathrm{Ni}_{62}\mathrm{Nb}_{38}$(~\cite{Zhang2016})                    &     1523 &    2.62(0.07) &    2.54(0.03) &      1.0(0.1) &      2.66(0.03) \\
$\mathrm{Pd}_{82}\mathrm{Si}_{18}$                                               &     1081 &    2.25(0.02) &    1.23(0.02) &     -2.2(0.3) &    2.472(0.002) \\
$\mathrm{Pt}$                                                                    &     2041 &    1.66(0.04) &      1.5(0.2) &     -2.1(0.9) &    1.656(0.006) \\
$^\dagger\mathrm{Pt}$(~\cite{Sheng2011})                                         &     1890 &    1.69(0.02) &  1.734(0.009) &  -0.344(0.01) &      1.54(0.03) \\
$^\dagger\mathrm{Si}$()                                                          &     1687 &    1.46(0.04) &     1.4(0.05) &    50.0(20.0) &      1.42(0.06) \\
$^\dagger\mathrm{Ta}$(~\cite{Zhong2014})                                         &     3290 &   0.858(0.03) &   0.755(0.03) &      0.2(0.1) &      1.03(0.02) \\
$\mathrm{Ti}$                                                                    &     1941 &     1.8(0.03) &    1.82(0.04) &      1.5(0.5) &  1.9151(0.0001) \\
$^\dagger\mathrm{Ti}$(~\cite{Mendelev2016})                                      &     1918 &    1.67(0.02) &    1.65(0.03) &    1.88(0.07) &      1.47(0.02) \\
$\mathrm{Ti}_{38.5}\mathrm{Zr}_{38.5}\mathrm{Ni}_{23}$                           &     1126 &    1.57(0.02) &    1.05(0.02) &    -0.94(0.4) &  1.7603(0.0008) \\
$\mathrm{Ti}_{40}\mathrm{Zr}_{10}\mathrm{Cu}_{30}\mathrm{Pd}_{20}$               &     1189 &    2.38(0.03) &    2.09(0.04) &    -0.62(0.5) &    2.514(0.002) \\
$\mathrm{Ti}_{40}\mathrm{Zr}_{10}\mathrm{Cu}_{36}\mathrm{Pd}_{14}$               &     1185 &    2.27(0.06) &    2.03(0.04) &   -0.072(0.2) &      2.52(0.08) \\
$\mathrm{Ti}_{45}\mathrm{Zr}_{45}\mathrm{Ni}_{10}$                               &     1543 &    1.66(0.09) &      1.2(0.1) &     -3.0(2.0) &    1.645(0.003) \\
$\mathrm{Vit105}$                                                                &     1093 &    1.57(0.03) &    0.94(0.04) &    -0.57(0.6) &    1.825(0.002) \\
$\mathrm{Vit106}$                                                                &     1123 &    1.53(0.02) &   0.841(0.04) &     -1.2(0.4) &    1.734(0.001) \\
$\mathrm{Vit106a}$                                                               &     1125 &    1.47(0.03) &   0.833(0.03) &    -0.29(0.3) &  1.7199(0.0007) \\
$\mathrm{Zr}$                                                                    &     2128 &    1.42(0.01) &    1.38(0.02) &    -0.15(0.8) &     1.52(0.003) \\
$^\dagger\mathrm{Zr}$(~\cite{Mendelev2007})                                      &     2109 &    1.92(0.07) &    1.67(0.04) &      1.1(0.1) &      1.79(0.04) \\
$\mathrm{Zr}_{50}\mathrm{Ti}_{50}$                                               &     1823 &    1.65(0.02) &    1.47(0.02) &     0.47(0.5) &    1.825(0.001) \\
$\mathrm{Zr}_{56}\mathrm{Co}_{28}\mathrm{Al}_{16}$                               &     1241 &    1.53(0.02) &   0.733(0.02) &     -1.8(0.5) &    1.797(0.001) \\
$\mathrm{Zr}_{57}\mathrm{Ni}_{43}$                                               &     1433 &    1.41(0.02) &    1.01(0.04) &    -0.91(0.6) &    1.731(0.001) \\
$\mathrm{Zr}_{59}\mathrm{Ti}_{3}\mathrm{Ni}_{8}\mathrm{Cu}_{20}\mathrm{Al}_{10}$ &     1145 &    1.47(0.02) &   0.825(0.02) &    -0.72(0.2) &    1.661(0.001) \\
$\mathrm{Zr}_{60}\mathrm{Ni}_{25}\mathrm{Al}_{15}$                               &     1248 &    1.36(0.06) &   0.612(0.04) &    -0.59(0.8) &    1.622(0.003) \\
$\mathrm{Zr}_{62}\mathrm{Ni}_{8}\mathrm{Cu}_{20}\mathrm{Al}_{10}$                &     1145 &    1.57(0.02) &   0.832(0.03) &     -1.1(0.2) &    1.777(0.001) \\
$\mathrm{Zr}_{64}\mathrm{Ni}_{25}\mathrm{Al}_{11}$                               &     1212 &    1.37(0.02) &   0.597(0.04) &    -0.66(0.6) &    1.635(0.001) \\
$\mathrm{Zr}_{65}\mathrm{Al}_{7.5}\mathrm{Cu}_{17.5}\mathrm{Ni}_{10}$            &     1170 &    1.45(0.02) &   0.763(0.02) &    -0.66(0.3) &    1.655(0.001) \\
$\mathrm{Zr}_{75.5}\mathrm{Pd}_{24.5}$                                           &     1303 &  1.417(0.009) &    1.04(0.02) &    -0.59(0.5) &   1.574(0.0008) \\
$\mathrm{Zr}_{80}\mathrm{Pt}_{20}$                                               &     1450 &    1.39(0.02) &    1.17(0.02) &     0.23(0.3) &  1.3692(0.0005) \\
$^\dagger\mathrm{Zr}_{80}\mathrm{Pt}_{20}$(~\cite{Hirata2013})                   &     1450 &    1.58(0.03) &    1.41(0.05) &      1.8(0.2) &      1.64(0.02) \\
$\mathrm{Zr}_{82}\mathrm{Ir}_{18}$                                               &     1513 &    1.36(0.02) &    1.09(0.03) &     -0.8(0.6) &    1.424(0.001) \\
\hline
\end{tabular}
\end{ruledtabular}
\end{table*}

\section{Conclusion}\label{conclusion}
The primary result of this study shows that to understand the change in the bond length as a function of temperature the nearest-neighbor distance distribution and robust measures of central tendency (such as the mean or median) should be considered, rather than $g(r)$ or any of its various forms that have typically been used. It is also shown that in agreement with recent studies~\cite{Ding2014,Ding2017a,Gangopadhyay2018} failing to account for the asymmetry of the NND distribution, but instead tracking the mode of the distribution, is the reason for the previously reported~\cite{Lou2013,Gangopadhyay2014d} anomalous contraction. The thermal expansion coefficient is shown to be directly related to the shift in the mean of the NND distribution, and that the rate of expansion in the bulk is likely equal to the expansion in the NND. This calls into question the methods used in a recent study that correlates the shift in the peak positions and volumetric dilation with the kinetic fragility~\cite{Wei2015}.

\section*{Acknowledgments}
This research used resources of the Advanced Photon Source, a U.S. Department of Energy (DOE) Office of Science User Facility operated for the DOE Office of Science by Argonne National Laboratory under Contract No. DEAC02-06CH11357. The work at Washington University in St. Louis was partially supported by the National Science Foundation under Grant DMR-12-06707 and the National Aeronautics Space Administration (NASA) under grant NNX10AU19G.

\bibliographystyle{apsrev4-1}
\bibliography{library}

\end{document}